\documentclass[graybox]{svmult}

\usepackage{helvet}         % selects Helvetica as sans-serif font
\usepackage{courier}        % selects Courier as typewriter font
\usepackage{type1cm}        % activate if the above 3 fonts are
\usepackage{makeidx}         % allows index generation
\usepackage{graphicx}        % standard LaTeX graphics tool
\usepackage{multicol}        % used for the two-column index
\usepackage[bottom]{footmisc}% places footnotes at page bottom

\usepackage{amsfonts}
\usepackage[usenames,dvipsnames]{xcolor}
\usepackage{amssymb,amsmath}

\def\nonu{\nonumber}

\def\br{\begin{eqnarray}}
\def\er{\end{eqnarray}} 
\def\be{\begin{equation}}
\def\bdm{\begin{displaymath}}
\def\ee{\end{equation}}
\def\edm{\end{displaymath}}
\def\0{\nonumber}

\def\({\left(}
\def\){\right)}

\relax

\newcommand{\bi}[1]{\bibitem{#1}}
\def\a{\alpha}
\def\b{\beta}

\def\d{\delta}

\def\g{k}

\def\l{\lambda}

\def\o{\over}

\def\pa{\partial}

\def\lie{{\cal G}}

\def\rlx{\relax\leavevmode}
\def\inbar{\vrule height1.5ex width.4pt depth0pt}
\def\IZ{\rlx\hbox{\sf Z\kern-.4em Z}}
\def\IR{\rlx\hbox{\rm I\kern-.18em R}}
\def\IC{\rlx\hbox{\,$\inbar\kern-.3em{\rm C}$}}
\def\one{\hbox{{1}\kern-.25em\hbox{l}}}

\def\nonu{\nonumber}
\def\br{\begin{eqnarray}}
\def\er{\end{eqnarray}}

\def\({\left(}
\def\){\right)}
\def\[{\left[}
\def\]{\right]}

\def\a{\alpha}
\def\b{\beta}

\def\d{\delta}

\def\bpsi{\bar{\psi}}

\def\g{\gamma}

\def\l{\lambda}

\def\o{\omega}

\def\pa{\partial}

\def\pt{\partial}

\def\tp0{\Theta_{+}^{(0)}}
\def\tm0{\Theta_{-}^{(0)}}

\def\bp{{\bar \p}}

\def\tp{\tilde{\phi}}

\def\bp{\bar{\psi}}
\def\bpp{\bar{\psi}_+}
\def\bpm{\bar{\psi}_-}
\def\dx{\pt_x\phi}
\def\dx{\phi_x}
\def\dxx{\phi_{2x}}
\def\dxxx{\phi_{3x}}

\def\dxm{\phi_{x}^{(-)}}
\def\dxxm{\phi_{2x}^{(-)}}
\def\dxxxm{\phi_{3x}^{(-)}}
\def\dxxxxm{\phi_{4x}^{(-)}}
\def\dxxxxxm{\phi_{5x}^{(-)}}
\def\dxp{\phi_{x}^{(+)}}
\def\dxxp{\phi_{2x}^{(+)}}
\def\dxxxp{\phi_{3x}^{(+)}}
\def\dxxxxp{\phi_{4x}^{(+)}}

\def\dxbpm{\bar{\psi}_{x}^{(-)}}
\def\dxxbpm{\bar{\psi}_{2x}^{(-)}}
\def\dxxxbpm{\bar{\psi}_{3x}^{(-)}}

\def\dxxxxxbpm{\bar{\psi}_{5x}^{(-)}}
\def\dxbpp{\bar{\psi}_{x}^{(+)}}
\def\dxxbpp{\bar{\psi}_{2x}^{(+)}}
\def\dxxxbpp{\bar{\psi}_{3x}^{(+)}}

\def\l{\lambda}

\def\nonu{\nonumber}

\def\bi{\begin{itemize}}
\def\ei{\end{itemize}}

\begin{document}

\title*{Recursion Operator and B\"acklund Transformation for Super mKdV Hierarchy}
\titlerunning{Recursion Operator and B\"acklund Transformation for smKdV Hierarchy} %for an abbreviated version of
\author{A.R. Aguirre, J.F. Gomes, A.L. Retore, N.I. Spano, and   A.H. Zimerman}
\institute{A.R. Aguirre \at Instituto de F\'isica e Qu\'imica, Universidade Federal de Itajub\'a - IFQ/UNIFEI, Av. BPS 1303,  37500-903, Itajub\'a, MG, Brasil. \email{alexis.roaaguirre@unifei.edu.br}
\and A.L. Retore \at Physics Department of the University of Miami, Coral Gables, FL 33124 USA. \email{retore@ift.unesp.br}
\and N.I. Spano, J.F. Gomes,  and A.H. Zimerman  \at Instituto de F\'isica Te\'orica - IFT/UNESP, Rua Dr. Bento Teobaldo Ferraz 271, Bloco II, 01140-070, S\~ao Paulo, Brasil. \email{jfg@ift.unesp.br, natyspano@ift.unesp.br, zimerman@ift.unesp.br}}

\maketitle

\abstract{In this paper we consider the  ${\cal N}=1$ supersymmetric  mKdV hierarchy composed of positive  odd flows  embedded  within an affine $\hat sl(2,1)$ algebra.
Its  B\"acklund transformations are constructed  in terms of a gauge transformation preserving the zero curvature representation.
The recursion operator  relating  consecutive flows is  derived   and   shown to  relate their Backlund transformations.  }

\section{Introduction}
\label{sec:1}

The algebraic formulation for integrable hierarchies  presents itself  as a powerful  framework  in order to discuss  its integrable properties, symmetries and soliton solutions.  
In particular the supersymmetric  mKdV  hierarchy  consists of a set of  time  evolution  (flows) equations  obtained from a zero curvature representation involving  a two dimensional gauge potential lying within an 
affine $\hat sl(2,1)$ Kac-Moody algebra and a common  infinite set of conservation laws \cite{1, Aguirre, Aratyn}.

Moreover,  B\"acklund transformation can be employed  to construct  an infinite sequence  of  solitons solutions 
by purely superposition principle  and also to link   nonlinear  equations to canonical  forms  as  discussed  for  many examples in \cite{Rogers}.
For the supersymmetric mKdV hierarchy,  the B\"acklund   transformation was derived   for the entire hierarchy by an universal  
gauge transformation  that preserves the zero curvature  representation and henceforth the equations of motion \cite{1}. 
The results obtained  in \cite{ Aguirre,  Ale3, Nathaly} for the super sinh-Gordon  were extended to the entire  smKdV hierarchy by the construction of a  B\"acklund-gauge transformation which  
connects  two field configurations of the same equations of motion \cite{1}.  
  Such structure  was first introduced  in \cite{Corr1,Corr2} for the bosonic sine-Gordon theory in order to 
  describe  integrable defects in the sense that two solitons solutions   are interpolated by a defect, as a set of internal boundary 
conditions derived from a Lagrangian density located at certain spatial position connecting two types of solutions. 
  %%%%%%%%%%%%%%%%%%%%%%%%%%%%%%%%%%%%%%%%%%%%%%%%%%%%%%%%%%%%%%%%%%%%%%%%%%
   The integrability  of the  model is  guaranteed  by the gauge invariance of the zero curvature representation.

 The  ${\cal N}=1$ supersymmetric modified Korteweg de-Vries (smKdV) hierarchy in the presence of defects    was  investigated in \cite{1} through the construction of 
gauge transformation in the form of  a B\"acklund-defect matrix approach. 
Firstly, we employ the defect matrix associated to the hierarchy which turns out  to be the same as  for the super sinh-Gordon (sshG) model.
The method is general  for all flows and   as an example  we have   derived 
  explicitly 
the B\"acklund equations in components for the  first few flows  of the hierarchy, namely $t_1, t_3$ and $t_5$. 
  Finally, this super B\"acklund transformation is employed to introduce
 type I defects for 
the supersymmetric mKdV 
hierarchy.  % Here  we  propose Further   integrability aspects by considering modified conserved quantities  are derived from the defect matrix.

  %%%%%%%%%%%%%%%%%%%%%%%%%%%%%%%%%%%%%%%%%%%%%%%%%%%%%%%

In this note  we propose an alternative derivation for  the B\"acklund  transformation  obtained  in \cite{1} by  employing  a recursion operator. 
For the bosonic  case of the mKdV hierarchy the recursion operator  was  constructed in \cite{Olver} and it relates equations  of motion for two consecutive  time evolutions. 
We show that the same philosophy  can be applied  to the supersymmetric mKdV hierarchy to relate   B\"acklund transformations  for two consecutive flows.

In what follows, we first  derive  the recursion operator for the  supersymmetric mKdV  hierarchy directly from the zero curvature representation. 
For technical reasons we change  variables $u(x, t_N)$ of  mKdV equation as  $ u(x, t_N) = \pa_x \phi(x, t_N)$  which seems more suitable to deal with B\"acklund transformations. 
We next conjecture that the B\"acklund transformations for consecutive  flows are also related  by the same recursion operator.  In fact we verify our conjecture  for the first few flows 
 generated by $t_1$,  $t_5$ and $t_3$.
 
\section{The smKdV hierarchy}
\label{sec:2}
% Always give a unique label
% and use \ref{<label>} for cross-references
% and \cite{<label>} for bibliographic references
% use \sectionmark{}
% to alter or adjust the section heading in the running head

An integrable hierarchy can be obtained from the zero curvature condition
%The supersymmetric mKdV hierarchy can be obtained from the zero curvature condition,
\begin{equation}
\left[ \pt_x+A_{x},\pt_{t_N}+A_{t_N}\right]=0
\label{zerocurv}
\end{equation}
where, $A_x$ and $A_{t_N}$ are the Lax pair  lying into an affine Kac-Moody superalgebra $(\widehat{\mathcal{G}})$ and $t_N$ represents the time flow of an integrable equation.\

Another important key ingredient  to construct an integrable hierarchy is a grading operator $Q$ and a constant grade one element $E^{(1)}$ that  decomposes the affine superalgebra into the following subspaces
\begin{eqnarray}\label{grade}
\widehat{\cal G} =\oplus \,\widehat{\cal G}_m = {\cal K}(E) \oplus {\cal M}(E)
\end{eqnarray}
where $m$ is the degree of the subspace $\widehat{\cal G}_m$ according to $Q$, i.e., $[Q, \lie_m] = m \lie_m$, ${\cal K}(E)=\big\{x\in \widehat{\mathcal{G}}\,/[\,x,E^{(1)}]=0 \big\}$ is the kernel 
of $E^{(1)}$ and ${\cal M}(E)$ is its complement ( image).\

Now we can define the Lax pair as
\begin{eqnarray}
A_x&=&E^{(1)}+A_{0}+A_{1/2},\\
A_{t_N}&=&D_{N}^{(N)}+D_{N}^{(N-1/2)}+...+D_{N}^{(1/2)}+D_{N}^{(0)},
\end{eqnarray}
where $A_0\in \widehat{\cal G}_0\cap{\cal M}(E), \, A_{1/2}\in\widehat{\cal G}_{1/2}\cap {\cal M}(E)$ with their respective components defining  the bosonic and fermionic fields of the theory and $D_N^{(m)}\in\widehat{\cal G}_m $.\

The decomposition of the equation (\ref{zerocurv}) into graded subspaces yields the following system\
\begin{eqnarray}
&& \,\,(N+1): \hspace{0.5cm} \left[E^{(1)},D_{N}^{(N)}\right]=0,\nonumber\\
&& \left(N+1/2\right):\hspace{0.2cm}  \left[E^{(1)},D_{N}^{(N-1/2)}\right] +\left[A_{1/2},D_{N}^{(N)}\right]=0, \nonumber\\
&& (N):\hspace{1.4cm} \pt_xD_{N}^{N}+\left[A_0,D_{N}^{(N)}\right]+\left[E^{(1)},D_{N}^{(N-1)}\right]+\left[A_{1/2},D_{N}^{(N-1/2)}\right]=0,\nonumber\\
&& \hspace{2.5cm}\vdots{}  \nonumber\\
&& \, (1): \hspace{1.5cm} \pt_xD_{N}^{(1)}+\left[A_0, D_{N}^{(1)}\right]+\left[E^{(1)},D_{N}^{(0)}\right]+\left[A_{1/2},D_{N}^{(1/2)}\right]=0,\nonumber\\
&& \left(1/2\right): \hspace{1.2cm} \pt_xD_{N}^{(1/2)}+\left[A_0, D_{N}^{(1/2)}\right]+\left[A_{1/2},D_{N}^{(0)}\right]-\pt_{t_N}A_{1/2}=0,\nonumber\\
&& \,\,(0): \hspace{1.5cm} \pt_xD_{N}^{(0)}+\left[A_0, D_{N}^{(0)}\right]-\pt_{t_N}A_{0}=0.
\label{decomposition}
\end{eqnarray} 
%Due to (\ref{grade}) each $D_N$ can be decomposed into kernel and image, ie, from the first equation in (\ref{decomposition}) we get that $D_N^{(N)}\in {\cal K}(E)$, now substituting this result in the second equation \\

The set of equations (\ref{decomposition}) can be recursively solved  yielding the time evolution equations for the fields in $A_0$ and $A_{1/2}$ as the zero and $1/2$ grade components, respectively.\

In particular, the construction of the supersymmetric mKdV hierarchy is based on  the judicious construction    of an affine subalgebra of  $\widehat{\cal G}=\widehat{sl}(2,1)$, with the principal 
gradation operator $Q=2d+\frac{1}{2}h_1^{(0)}$ and the constant element $E^{(1)}=K^{(1)}+K^{(2)}$.  Its generators  may be  regrouped  as
\br
&& F_1^{\left(2n+\frac{3}{2}\right)}=E_{\a_1+\a_2}^{\left(n+\frac{1}{2}\right)}-E_{\a_2}^{(n+1)}+E_{-\a_1-\a_2}^{(n+1)}-E_{-\a_2}^{\left(n+\frac{1}{2}\right)},\nonu\\
&& F_2^{\left(2n+\frac{1}{2}\right)}=-E_{\a_1+\a_2}^{(n)}+E_{\a_2}^{\left(n+\frac{1}{2}\right)}+E_{-\a_1-\a_2}^{\left(n+\frac{1}{2}\right)}-E_{-\a_2}^{(n)},\nonu\\
&& G_1^{\left(2n+\frac{1}{2}\right)}=E_{\a_1+\a_2}^{(n)}+E_{\a_2}^{\left(n+\frac{1}{2}\right)}+E_{-\a_1-\a_2}^{\left(n+\frac{1}{2}\right)}+E_{-\a_2}^{(n)}\nonu
\er
\begin{eqnarray}
&& G_2^{\left(2n+\frac{3}{2}\right)}=-E_{\a_1+\a_2}^{\left(n+\frac{1}{2}\right)}-E_{\a_2}^{(n+1)}+E_{-\a_1-\a_2}^{(n+1)}+E_{-\a_2}^{\left(n+\frac{1}{2}\right)},\nonu\\
&& K_1^{(2n+1)}=-E_{-\a_1}^{(n+1)}-E_{\a_1}^{(n)},\nonu\\
&& K_2^{(2n+1)}=h_{1}^{\left(n+\frac{1}{2}\right)}-h_{2}^{\left(n+\frac{1}{2}\right)},\nonu\\
&& M_1^{(2n+1)}=E_{-\a_1}^{(n+1)}-E_{\a_1}^{(n)},\nonu\\
&& M_2^{(2n)}=2h_{1}^{(2n)}
\label{affine} 
\end{eqnarray}
and  decomposed as follows (see \cite{Ymai} for details),\
\begin{eqnarray}
&& {\cal M}_{bos}=\left\{M_2^{(2n)}, \,M_1^{(2n+1)}\right\},\qquad {\cal M}_{fer}=\left\{G_1^{(2n+\frac{1}{2})}, \, G_2^{(2n+\frac{3}{2})}\right\},\nonumber\\
&& {\cal K}_{bos}=\left\{K_1^{(2n+1)},\,K_2^{(2n+1)}\right\},\qquad {\cal K}_{fer}=\left\{F_1^{(2n+\frac{3}{2})},\,F_2^{(2n+\frac{1}{2})}\right\}
\label{grading}
\end{eqnarray}

Notice that the fermionic generators $F_i$ and $G_i, i=1,2$  lying in the Kernel and Image respectively  display  an explicit $Z_2$  structure in their affine indices, 
in the sense that the semi integers indices $N+ 1/2$  are decomposed  according to $2n+1/2$ and $2n+3/2$ disjoints subsets.  Another $Z_2$ structure arises, now decomposing the integers $N$   into odd $(2n+1)$  and even $(2n)$ subsets.  
Assign to the bosonic  generators  $\{ K_1,\, K_2,\, M_1 \}$  and  $\{ M_2 \}$  the grades $2n+1$ and $2n$  respectively.  
The  affine algebra displayed in the appendix is shown to close  consistently with the  $Z_2$  structures described above.\

The $x$ part of the Lax pair is then constructed  from  $A_0=u(x,t)M_2^{(0)}$ and $A_{1/2}=\bp(x,t)G_1^{(1/2)}$. \

The first equation in the system (\ref{decomposition}) implies  that $D_N^{(N)}\in {\cal K}(E)$ and hence $N=2n+1$. 
In order to solve equations in (\ref{decomposition}) we expand  $D_N^{(m)}$ according to  its  bosonic or fermionic  character  using latin or greek  coefficients, respectively  
following the grading given in equation (\ref{grading}) {\footnote{Moreover we use $\a_m, \b_m$ for grades $m=2n+1/2$ and $\g_m, \d_m$ for $m=2n+3/2$  while $a_m,b_m,c_m$ for $m=2n+1$ and $d_m$ for $m=2n$.}}, i.e.,
\br
&& D_{N}^{(2n+\frac{3}{2})}=\g_{2n+\frac{3}{2}}F_1^{\left(2n+\frac{3}{2}\right)}+\d_{2n+\frac{3}{2}}G_2^{\left(2n+\frac{3}{2}\right)},\nonu\\
&& D_{N}^{(2n+1)}={a}_{2n+1}K_1^{(2n+1)}+b_{2n+1}K_2^{(2n+1)}+c_{2n+1}M_1^{(2n+1)}, \nonumber\\
&& D_{N}^{(2n+\frac{1}{2})}=\a_{2n+\frac{1}{2}}F_2^{\left(2n+\frac{1}{2}\right)}+\b_{2n+\frac{1}{2}}G_1^{\left(2n+\frac{1}{2}\right)}, \nonumber\\
&& D_{N}^{(2n)}=d_{2n}M_2^{(2n)},\nonu\\
&& D_{N}^{(2n-\frac{1}{2})}=\g_{2n-\frac{1}{2}}F_1^{\left(2n-\frac{1}{2}\right)}+\d_{2n-\frac{1}{2}}G_2^{\left(2n-\frac{1}{2}\right)},\nonu\\
&& D_{N}^{(2n-1)}={a}_{2n-1}K_1^{(2n-1)}+b_{2n-1}K_2^{(2n-1)}+c_{2n-1}M_1^{(2n-1)}, \nonumber\\
&& D_{N}^{(2n-\frac{3}{2})}=\a_{2n-\frac{3}{2}}F_2^{\left(2n-\frac{3}{2}\right)}+\b_{2n-\frac{3}{2}}G_1^{\left(2n-\frac{3}{2}\right)}, \nonumber\\
&& D_{N}^{(2n-2)}=d_{2n-2}M_2^{(2n-2)},\nonu\\
&& \vdots \nonu\\
&& D_{N}^{(\frac{3}{2})}=\g_{\frac{3}{2}}F_1^{\left(\frac{3}{2}\right)}+\d_{\frac{3}{2}}G_2^{\left(\frac{3}{2}\right)},\nonu\\
&& D_{N}^{(1)}={a}_{1}K_1^{(1)}+b_{1}K_2^{(1)}+c_{1}M_1^{(1)}, \nonumber\\
&& D_{N}^{(\frac{1}{2})}=\a_{\frac{1}{2}}F_2^{\left(\frac{1}{2}\right)}+\b_{\frac{1}{2}}G_1^{\left(\frac{1}{2}\right)}\nonu\\
&& D_{N}^{(0)}=d_{0}M_2^{(0)}.
\label{Dn}
\end{eqnarray}
%\noindent 
where the $a_m, b_m, c_m, d_m$ and $\a_m, \b_m, \g_m, \d_m$ are functionals of the fields $u$ and $\bpsi$. \

Substituting this parameterization in the equation (\ref{decomposition}), one  solve recursively for all  $D^{(m)}, m=0,\cdots N$. Starting with the highest  grade equation in  (\ref{decomposition})  in which  $N=2n+1 $,\
\begin{equation}
[K_1^{(1)}+K_2^{(1)},{a}_{2n+1}K_1^{(2n+1)}+b_{2n+1}K_2^{(2n+1)}+c_{2n+1}M_1^{(2n+1)}]=0
\end{equation}

We obtain after using the comutation relations given  in the appendix that $c_{2n+1}=0$. Now substituting this result in the next equation in (\ref{decomposition}), i.e, the equation for degree $N+1/2$ we get,\
\begin{equation}
\b_{2n+\frac{1}{2}}=\frac{1}{2}\bp(a_{2n+1}+b_{2n+1})
\end{equation}

From the equation for degree $N$ we find that $a_{2n+1}, b_{2n+1}$ are constants and $d_{2n}=u a_{2n+1}+\bp \a_{2n+\frac{1}{2}}$. Proceeding in this way until we reach the equation for degree $N-2$, we get\
\begin{eqnarray}
(N-1/2): \hspace{0.5cm} &&\pt_x\a_{2n+\frac{1}{2}}-u \b_{2n+\frac{1}{2}}+\bp d_{2n}=0\nonu\\
&&\pt_x\b_{2n+\frac{1}{2}}-u \a_{2n+\frac{1}{2}}+2\d_{2n-\frac{1}{2}}=0\label{N-1/2}\\
(N-1): \hspace{0.5cm} &&\pt_x d_{2n}-2 c_{2n-1}+2\bp \g_{2n-\frac{1}{2}}=0\label{N-1}\\
(N-3/2): \hspace{0.5cm} &&\pt_x\g_{2n-\frac{1}{2}}-u \d_{2n-\frac{1}{2}}+\bp c_{2n-1}=0\nonu\\
&&\pt_x\d_{2n-\frac{1}{2}}-u \g_{2n-\frac{1}{2}}+2\b_{2n-\frac{3}{2}}-\bp(a_{2n-1}+b_{2n-1})=0\label{N-3/2}\\
(N-2): \hspace{0.5cm} &&\pt_x a_{2n-1}+2u c_{2n-1}-2\bp \b_{2n-\frac{3}{2}}=0\nonu\\
&&\pt_x b_{2n-1}+2\bp \b_{2n-\frac{3}{2}}=0\nonu\\
&&\pt_x c_{2n-1}-2 d_{2n-2}+2u a_{2n-1}+2\bp\a_{2n-\frac{3}{2}}=0\label{N-2}
\end{eqnarray}

The subsequent equations are all similar to the set above, in the sense that the equations for even grade will correspond to (\ref{N-1}), the odd ones will be similar to the set in (\ref{N-2}). \

For the semi-integer degree equations the following combinations are allowed: if $(N-\frac{1}{2}-2m)$ then it corresponds to the set (\ref{N-1/2}) and if the grade can be written as $(N-\frac{1}{2}-2m-1)$ it seems like (\ref{N-3/2}) where $m\in Z_+$.\

Then for a specific $n\in Z_+$ these results can be written in the following way,
\begin{eqnarray}\label{sistema}
&& c_{2n+1}=0,\qquad \beta_{2n+\frac{1}{2}}=\frac{\sqrt{i}}{2}\bp(a_{2n+1}+b_{2n+1}),\nonu\\
&& a_{2n+1}=constant\qquad b_{2n+1}=constant\nonu\\
&& d_{2n}=u a_{2n+1}+\sqrt{i}\bp \a_{2n+\frac{1}{2}}\nonu\\
&&\pt_x\a_{2n+\frac{3}{2}-j}-u \beta_{2n+\frac{3}{2}-j}+\sqrt{i}\bp d_{2n+1-j}=0\,\,\,\,\, \qquad \quad\qquad\qquad\quad\qquad\textit{(odd j)}\nonu\\
&&\pt_x\beta_{2n+\frac{3}{2}-j}-u \a_{2n+\frac{3}{2}-j}+2\d_{2n+\frac{1}{2}-j}=0\, \,\qquad\qquad \qquad\qquad\quad\qquad\quad\textit{(odd j)}\nonu\\
&& \pt_x d_{2n+1-j}-2c_{2n-j}+2\sqrt{i}\bp \g_{2n+\frac{1}{2}-j}=0\,\,\,\,\,\, \qquad\qquad\qquad\qquad \qquad\quad\textit{(odd j)}\nonu\\
&&\pt_x\g_{2n+\frac{3}{2}-j}-u \d_{2n+\frac{3}{2}-j}+\sqrt{i}\bp c_{2n+1-j}=0\,\,\,\,\,\,\, \, \qquad \qquad\qquad\qquad\qquad\textit{(even j)}\nonu\\
&&\pt_x\d_{2n+\frac{3}{2}-j}-u \g_{2n+\frac{3}{2}-j}+2\beta_{2n+\frac{1}{2}-j}-\sqrt{i}\bp(a_{2n+1-j}+b_{2n+1-j})=0\,  \textit{(even j)}\nonu\\
&& \pt_x a_{2n+1-j}+2u c_{2n+1-j}-2\sqrt{i}\bp \beta_{2n+\frac{1}{2}-j}=0\,\,\,\,\,\,\ \quad\quad\qquad\qquad\qquad\quad \textit{(even j)}\nonu\\
&&\pt_x b_{2n+1-j}+2\sqrt{i}\bp \beta_{2n+\frac{1}{2}-j}=0\,\,\,\,\,\,\,\qquad\qquad\qquad\qquad\qquad\qquad \qquad\quad\textit{(even j)}\nonu\\
&&\pt_x c_{2n+1-j}-2 d_{2n-j}+2u a_{2n+1-j}+2\sqrt{i}\bp\a_{2n+\frac{1}{2}-j}=0 \qquad\qquad\quad \quad\textit{(even j)}\nonu\\
\end{eqnarray}
where $j=1,...,2n$. \

We proceed in this way until the grade  $(1/2)$ equation in (\ref{decomposition}) to get
\begin{eqnarray}
\pt_x\a_{\frac{1}{2}}&=&u\b_{\frac{1}{2}}-\bp d_0\\
\pt_{t_{2n+1}}\bp&=&\pt_x\b_{\frac{1}{2}}-u\a_{\frac{1}{2}}\label{t2n+1psi}
\end{eqnarray}
and the zero  grade equation  to obtain \
\begin{equation}
\pt_{t_{2n+1}}u=\pt_x d_0\label{t2n+1u}
\end{equation} 

Therefore the problem is to recursively solve  this set of equations, finding the respective coefficients  for a given  value of $n$ and then substitute them 
in (\ref{t2n+1u}) and (\ref{t2n+1psi}) to obtain the time evolution of the fields $u$, $\bp$. For example, if $n=0$ the equations of motion are\
\begin{equation}
\pt_{t_{1}}\bp=\pt_x\bp,\qquad \pt_{t_{1}}u=\pt_x u\label{t1}
\end{equation} 

For $n=1$ we have the supersymmetric mKdV equation\
\begin{eqnarray}
&& 4\pt_{t_3}u=u_{3x}-6u^2u_x+3i\bp\pt_x\left(u\bp_x\right),\label{movu}\\
&& 4\pt_{t_3}\bp=\bp_{3x}-3u\pt_x\left(u\bp\right).\label{movpsi}
\end{eqnarray} 
%\newpage

Then, for $n=2$ we have
\br
16\pt_{t_5}u &=&u_{5x}-10(u_x)^3-40 u(u_x)(u_{2x})-10 u^2(u_{3x})+30 u^4 (u_x)\nonu\\[0.2cm]
&+&5 i \bp\pt_x(u \bp_{3x}-4 u^3 \bp_x+u_x\bp_{2x}+u_{2x} \bp_x)+5 i \bp_x\pt_x(u\bp_{2x}),\label{t5u}\\[0.2cm]
16\pt_{t_5}\bp &=&\bp_{5x}-5 u\pt_x(u\bp_{2x}+2u_x\bp_x+u_{2x}\bp)+10 u^2\pt_x(u^2\bp)\nonu\\&-&10(u_x)\pt_x(u_x\bp).\label{t5psi}
\er

\section{Recursion operator for smKdV hierarchy} 
We shall now consider the construction of a set of supersymmetric  integrable equations by solving the system in (\ref{sistema}). Since the solution of (\ref{sistema}) is similar for all values of  $n$ 
it is expected that there exists a connection among  the time flows. The recursion operator is the mathematical object responsible for such connection and  will be constructed in this section. \

In order to see this we consider the equations for $N=2n+1$ and $N=2n+3$\
\begin{eqnarray}
&&{\bf 2n+1}\qquad\quad \quad\qquad\qquad\qquad\qquad{\bf 2n+3}\nonu\\
&& c_{2n+1}=0,\qquad\quad\quad\qquad\qquad\qquad\,\, \, \, c_{2n+3}=0\\
&& a_{2n+1}=b_{2n+1}=1,\,\quad\quad\quad\qquad \qquad a_{2n+3}=b_{2n+3}=1,\\
&& \b_{2n+\frac{1}{2}}=\bp \quad\qquad\qquad\qquad\qquad\qquad \b_{2n+\frac{5}{2}}=\bp\\
&& d_{2n}=u +\bp \a_{2n+\frac{1}{2}},\,\quad\qquad\qquad\qquad d_{2n+2}=u +\bp \a_{2n+\frac{5}{2}}\\
&& \pt_x\a_{2n+\frac{1}{2}}-u \b_{2n+\frac{1}{2}}+\bp d_{2n}=0,  \,\,\, \quad  \pt_x\a_{2n+\frac{5}{2}}-u \b_{2n+\frac{5}{2}}+\bp d_{2n+2}=0\\
&& \pt_x\b_{2n+\frac{1}{2}}-u \a_{2n+\frac{1}{2}}+2\d_{2n-\frac{1}{2}}=0 ,\, \, \, \pt_x\b_{2n+\frac{5}{2}}-u \a_{2n+\frac{5}{2}}+2\d_{2n+\frac{3}{2}}=0\\
&&\pt_x d_{2n}-2c_{2n-1}+2\bp \g_{2n-\frac{1}{2}}=0,\, \, \quad \pt_x d_{2n+2}-2c_{2n+1}+2\bp \g_{2n+\frac{3}{2}}=0\\
&&\vdots\qquad\qquad\qquad\qquad\qquad\qquad\qquad\qquad\vdots \nonu\\
&&\pt_x\a_{1/2}-u\b_{1/2}+\bp d_0=0,\quad\quad\qquad\pt_x\a_{5/2}-u\b_{5/2}+\bp d_2=0\\
&&\, \, \quad\qquad\qquad\qquad\qquad\qquad\qquad\qquad\, \,\pt_x\b_{5/2}-u\a_{5/2}+2\d_{3/2}=0\label{eq1}\\
&&\quad\qquad\qquad\qquad\,\qquad\qquad\qquad\qquad\, \, \pt_x d_{2}-2 c_{1}+2\bp\g_{3/2}=0\label{eq2}\\
&&\, \, \quad\qquad\qquad\qquad\qquad\qquad\qquad\qquad\pt_x\g_{3/2}-u\d_{3/2}+\bp c_{1}=0\label{eq3}\\
&&\,\, \quad\qquad\qquad\qquad\qquad\qquad\qquad\qquad\pt_x\d_{3/2}-u\g_{3/2}-\bp(a_1+b_1)+2\b_{1/2}=0\nonu\\ \label{eq4}\\
&&\qquad\qquad\qquad\qquad\qquad\qquad\qquad\qquad\pt_x a_1-2\bp\b_{1/2}+2u c_1=0\label{eq5}\\
&&\qquad\qquad\qquad\qquad\qquad\qquad\qquad\qquad\pt_x b_1+2\bp \b_{1/2}=0\label{eq6}\\
&&\qquad\qquad\qquad\qquad\qquad\qquad\qquad\qquad\pt_x c_1-2d_0+2u a_1+2\bp \a_{1/2}=0\label{eq7}\\
&&\qquad\qquad\qquad\qquad\qquad\qquad\qquad\qquad\pt_x\a_{1/2}-u\b_{1/2}+\bp d_0=0\label{eq8}\\
&&\pt_{t_{2n+1}}\bp=\pt_x \b_{1/2}-u\a_{1/2},\quad\qquad\quad\qquad\pt_{t_{2n+3}}\bp=\pt_x\b_{1/2}-u\a_{1/2}\label{movp5}\\
&&\pt_{t_{2n+1}}u=\pt_x d_0,\quad\,\, \qquad \qquad\qquad\quad\qquad\pt_{t_{2n+3}}u=\pt_x d_0\label{movu5}
\end{eqnarray}
%\end{multicols}
%The last two equations provide the equation of motion for the smKdV.\\

Notice that until the equation (\ref{eq1}) the alligned equations have the same solution, in such a way that we can make the following useful identifications, 
\begin{equation}\label{id}
d_2\Big|_{2n+3}=d_0\Big|_{2n+1},\quad \beta_{5/2}\Big|_{2n+3}=\beta_{1/2}\Big|_{2n+1},\quad \a_{5/2}\Big|_{2n+3}=\a_{1/2}\Big|_{2n+1}.
\end{equation}

The case for $N=2n+3$ has eight additional equations (\ref{eq1})-(\ref{eq8}), which can be solved in terms of the coefficients for $N=2n+1$ by the relations (\ref{id}). Then we will be able to relate the time evolution equations for $t_{2n+3}$ to the time evolution equations for $t_{2n+1}$. \

Starting with the equation (\ref{eq1}) by using (\ref{id}) we get
\begin{equation}
\d_{3/2}\Big|_{2n+3}=-\frac{1}{2}\pt_{t_{2n+1}}\bp.
\end{equation}

Substituting in (\ref{eq2}) and (\ref{eq3})
\begin{eqnarray}
c_1\Big|_{2n+3}&=&\frac{1}{2}\pt_{t_{2n+1}}u+\bp\g_{3/2}\Big|_{2n+3}\\
\g_{3/2}\Big|_{2n+3}&=&-\frac{1}{2}\int dx\,\pt_{t_{2n+1}}(u\bp).
\end{eqnarray}\

Recursively solving  the equations (\ref{eq4})-(\ref{eq8}) we get the following coefficients
\begin{eqnarray}
\b_{1/2}\Big|_{2n+3}&=&\frac{1}{4}\pt_x\pt_{t_{2n+1}}\bp-\frac{u}{4}\int dx\,\pt_{t_{2n+1}}(u\bp)+\frac{1}{2}\bp(a_1+b_1)\Big|_{2n+3},\\
a_1\Big|_{2n+3}&=&-\int dx\, u\pt_{t_{2n+1}}u+\frac{1}{2}\int dx'\, u\bp \int dx\, \pt_{t_{2n+1}}(u\bp)\nonu\\&+&\frac{1}{2}\int dx\, \bp\pt_x\pt_{t_{2n+1}}\bp ,\\
b_1\Big|_{2n+3}&=&-\frac{1}{2}\int dx\, \bp\pt_x\pt_{t_{2n+1}}\bp+\frac{1}{2}\int dx'\, u\bp \int dx\, \pt_{t_{2n+1}}(u\bp),\\
d_0\Big|_{2n+3}&=&\frac{1}{4}\pt_x\pt_{t_{2n+1}}u-\frac{1}{4}\bp\pt_{t_{2n+1}}(u\bp)-u\int dx\, u\pt_{t_{2n+1}}u+\bp\a_{1/2}\Big|_{2n+3}\nonu\\
&+&\frac{u}{2}\int dx'\, u\bp \int dx\, \pt_{t_{2n+1}}(u\bp)-\frac{\pt_x\bp}{4}\int dx\, \pt_{t_{2n+1}}(u\bp)\nonu\\ 
&+&\frac{u}{2}\int dx\, \bp\pt_x\pt_{t_{2n+1}}\bp ,\\
\a_{1/2}\Big|_{2n+3}&=&\frac{1}{4}\int dx\, (u\pt_x\pt_{t_{2n+1}}\bp-\bp\pt_x\pt_{t_{2n+1}}u)\nonu\\
&+&\frac{1}{2}\int dx'\, u\bp \int dx\, (u\pt_{t_{2n+1}}u-\bp\pt_x\pt_{t_{2n+1}}\bp)\nonu\\
&-&\frac{1}{4}\int dx'\, \left(u^2-\bp\pt_x\bp\right) \int dx\, \pt_{t_{2n+1}}(u\bp).
\end{eqnarray}

Finally putting these coefficients in the equations of motion (\ref{movp5}) and (\ref{movu5}) we obtain that the $t_{2n+3} $ equation of the smKdV hierarchy is given by
\begin{equation}
\frac{\pt u}{\pt t_{2n+3}}=R_1\frac{\pt u}{\pt t_{2n+1}}+R_2\frac{\pt \bp}{\pt t_{2n+1}},\qquad \frac{\pt \bp}{\pt t_{2n+3}}=R_3\frac{\pt u}{\pt t_{2n+1}}+R_4\frac{\pt \bp}{\pt t_{2n+1}}
\label{rec}
\end{equation}
where $ \{R_1, R_4\}$, $\{R_2, R_3\}$ are the bosonic and fermionic components of the recursion operator, respectively,  which are given by
\br
R_1&=&\frac{1}{4}\mathbb{D}^2-u^2-u_x\mathbb{D}^{-1}u+\frac{i}{4}\bp\bp_x+\frac{i}{4}u^2\bp \mathbb{D}^{-1}\bp+\frac{i}{2}u_x \mathbb{D}^{-1}u\bp \mathbb{D}^{-1}\bp\nonu\\[0,1cm]
&-&-\frac{i}{4}\bp_{2x} \mathbb{D}^{-1}\bp-\frac{i}{4}\bp_x \mathbb{D}^{-1}\bp \mathbb{D}-\frac{i}{4}\bp_x \mathbb{D}^{-1}u^2 \mathbb{D}^{-1}\bp+\frac{i}{2}\bp_x \mathbb{D}^{-1}u\bp \mathbb{D}^{-1}u\nonu\\[0,1cm]
&-&\frac{1}{4}\bp_x \mathbb{D}^{-1}\bp\bp_x \mathbb{D}^{-1}\bp,%\\[0,2cm]
\er

\begin{eqnarray}
R_2&=&\frac{i}{2}u\bp \mathbb{D}-\frac{i}{2}u\bp_x-\frac{i}{4}u_x\bp+\frac{i}{4}u^2\bp \mathbb{D}^{-1}u+\frac{i}{2}u_x \mathbb{D}^{-1}\bp \mathbb{D}+\frac{i}{2}u_x \mathbb{D}^{-1}u\bp \mathbb{D}^{-1}u\nonu\\[0.1cm]
&-&\frac{i}{4}\bp_{2x} \mathbb{D}^{-1}u
+\frac{i}{4}\bp_x \mathbb{D}^{-1}u \mathbb{D}-\frac{i}{4}\bp_x \mathbb{D}^{-1}u^2 \mathbb{D}^{-1}u+\frac{1}{2}\bp_x \mathbb{D}^{-1}u\bp \mathbb{D}^{-1}\bp \mathbb{D}\nonu\\[0.1cm]
&-&\frac{1}{4}\bp_x \mathbb{D}^{-1}\bp\bp_x \mathbb{D}^{-1}u,\\[0,2cm]
R_3&=&-\frac{3}{4}u\bp-\frac{1}{4}u_x \mathbb{D}^{-1}\bp-\frac{1}{2}\bp_x\mathbb{D}^{-1}u+\frac{i}{2}\bp_x \mathbb{D}^{-1}u\bp \mathbb{D}^{-1}\bp+\frac{1}{4}u \mathbb{D}^{-1}\bp \mathbb{D}\nonu\\[0,1cm]
&-&\frac{1}{2}u \mathbb{D}^{-1}u\bp \mathbb{D}^{-1}u-\frac{i}{4}u \mathbb{D}^{-1}\bp\bp_x \mathbb{D}^{-1}\bp+\frac{1}{4}u \mathbb{D}^{-1}u^2 \mathbb{D}^{-1}\bp,\\[0,2cm]
R_4&=&\frac{1}{4}\mathbb{D}^2-\frac{1}{4}u^2-\frac{1}{4}u_x\mathbb{D}^{-1}u-\frac{1}{4}u \mathbb{D}^{-1}u \mathbb{D}+\frac{1}{4}u \mathbb{D}^{-1}u^2 \mathbb{D}^{-1}u\nonu\\[0,1cm]
&-&\frac{i}{4}u \mathbb{D}^{-1}\bp\bp_x \mathbb{D}^{-1}u+\frac{i}{2}\bp_x \mathbb{D}^{-1}u\bp \mathbb{D}^{-1}u+\frac{i}{2}u \mathbb{D}^{-1}u\bp \mathbb{D}^{-1}\bp \mathbb{D}.
\end{eqnarray}
where $\mathbb{D}=\pt_x$ and $\mathbb{D}^{-1}$ is its inverse. \ %In terms of $u=-\pt_x\phi$ we get

In terms of $u=\dx$ we get 
\begin{equation}\label{recursionphi}
\frac{\pt \phi}{\pt t_{2n+3}}=\mathbb{R}_1\frac{\pt \phi}{\pt t_{2n+1}}+\mathbb{R}_2\frac{\pt \bp}{\pt t_{2n+1}},\qquad
\frac{\pt \bp}{\pt t_{2n+3}}=\mathbb{R}_3\frac{\pt \phi}{\pt t_{2n+1}}+\mathbb{R}_4\frac{\pt \bp}{\pt t_{2n+1}}
\end{equation}
where $\mathbb{R}_1=\mathbb{D}^{-1}\mathcal{R}_1 \mathbb{D},\, \mathbb{R}_2=\mathbb{D}^{-1}\mathcal{R}_2,\, \mathbb{R}_3=\mathcal{R}_3 \mathbb{D},\, \mathbb{R}_4=\mathcal{R}_4 $, with\
\br
\mathcal{R}_1&=&\frac{1}{4}\mathbb{D}^2-\dx^2-\dxx \mathbb{D}^{-1}\dx+\frac{i}{4}\bp\bp_x+\frac{i}{4}\dx^2\bp \mathbb{D}^{-1}\bp-\frac{i}{4}\bp_{2x} \mathbb{D}^{-1}\bp\nonu\\[0,1cm]
&-&\frac{i}{4}\bp_x \mathbb{D}^{-1}\bp \mathbb{D}-\frac{i}{4}\bp_x \mathbb{D}^{-1}\dx^2 \mathbb{D}^{-1}\bp+\frac{i}{2}\bp_x \mathbb{D}^{-1}\dx\bp \mathbb{D}^{-1}\dx\nonu\\[0,1cm]
&-&\frac{1}{4}\bp_x \mathbb{D}^{-1}\bp\bp_x \mathbb{D}^{-1}\bp+\frac{i}{2}\dxx \mathbb{D}^{-1}\dx\bp \mathbb{D}^{-1}\bp,\\[0.2cm]
\mathcal{R}_2&=&\frac{i}{2}\dx\bp \mathbb{D}-\frac{i}{2}\dx\bp_x-\frac{i}{4}\dxx\bp+\frac{i}{4}\dx^2\bp \mathbb{D}^{-1}\dx+\frac{i}{2}\dxx \mathbb{D}^{-1}\bp \mathbb{D}-\frac{i}{4}\bp_{2x} \mathbb{D}^{-1}\dx\nonu\\[0.1cm]
&+&\frac{i}{4}\bp_x \mathbb{D}^{-1}\dx \mathbb{D}-\frac{i}{4}\bp_x \mathbb{D}^{-1}\dx^2 \mathbb{D}^{-1}\dx+\frac{1}{2}\bp_x \mathbb{D}^{-1}\dx\bp \mathbb{D}^{-1}\bp \mathbb{D}\nonu\\[0.1cm]
&-&\frac{1}{4}\bp_x \mathbb{D}^{-1}\bp\bp_x \mathbb{D}^{-1}\dx+\frac{i}{2}\dxx \mathbb{D}^{-1}\dx\bp \mathbb{D}^{-1}\dx,\\[0.2cm]
\mathcal{R}_3&=&-\frac{3}{4}\dx\bp-\frac{1}{4}\dxx \mathbb{D}^{-1}\bp-\frac{1}{2}\bp_x\mathbb{D}^{-1}\dx+\frac{i}{2}\bp_x \mathbb{D}^{-1}\dx\bp \mathbb{D}^{-1}\bp+\frac{1}{4}\dx \mathbb{D}^{-1}\bp \mathbb{D}\nonu\\[0.1cm]
&-&\frac{1}{2}\dx \mathbb{D}^{-1}\dx\bp \mathbb{D}^{-1}\dx-\frac{i}{4}\dx \mathbb{D}^{-1}\bp\bp_x \mathbb{D}^{-1}\bp+\frac{1}{4}\dx \mathbb{D}^{-1}\dx^2 \mathbb{D}^{-1}\bp,%\\[0.2cm]
\er
\begin{eqnarray}
\mathcal{R}_4&=&\frac{1}{4}\mathbb{D}^2-\frac{1}{4}\dx^2-\frac{1}{4}\dxx \mathbb{D}^{-1}\dx-\frac{1}{4}\dx \mathbb{D}^{-1}\dx \mathbb{D}+\frac{1}{4}\dx \mathbb{D}^{-1}\dx^2 \mathbb{D}^{-1}\dx\nonu\\[0.1cm]
&-&\frac{i}{4}\dx \mathbb{D}^{-1}\bp\bp_x \mathbb{D}^{-1}\dx+\frac{i}{2}\bp_x \mathbb{D}^{-1}\dx\bp \mathbb{D}^{-1}\dx+\frac{i}{2}\dx \mathbb{D}^{-1}\dx\bp \mathbb{D}^{-1}\bp \mathbb{D}.
\end{eqnarray}

We  have explicitly checked that by  employing equation (\ref{rec}) for $n=0$ we  recover the smKdV equation (\ref{movu}), (\ref{movpsi}).  Also  it was  verified that  (\ref{rec}) for   $n=1$, yields  the $t_{5}$ flow of  the hierarchy (\ref{t5u}), (\ref{t5psi}) as predicted.

\section{The B\"acklund transformations for the smKdV hierarchy }
In this section we will start by reviewing the systematic construction of the B\"acklund transformation for the smKdV hierarchy, based on the invariance of the zero curvature equation (\ref{zerocurv}) under the gauge transformation,
\begin{equation}\label{gauge}
\pt_\mu K=KA_\mu(\phi_1,\bar{\psi}_1)-A_\mu(\phi_2,\bar{\psi}_2)K. 
\end{equation}
where $ A_{1}= A_x,\quad A_0 =A_{t_{2n+1}}, \quad \pa_1 = \pa_x, \quad \pa_0 = \pa_{t_{2n+1}}$   and assuming the existence of a defect matrix  $K(\phi_1,\bp_1,  \phi_2,\bp_2  ) $ which maps a field configuration $\{\phi_1,\bp_1\}$ into another $\{\phi_2,\bp_2\}$.\

%and it will also provide the corresponding B\"acklund transformations. \\
It is important to point out that the spatial Lax operator $A_x$ is common to all members of the smKdV hierarchy and is given by  
\begin{equation}
\renewcommand{\arraystretch}{2}
A_x=\left(
\begin{array}{cc|c}
\lambda^{1/2}-\dx & -1 & \sqrt{i}\,\bpsi \hspace{0.2cm}\\
-\lambda & \lambda^{1/2}+\dx & \sqrt{i}\,\lambda^{1/2}\,\bpsi\\\hline
\sqrt{i}\,\lambda^{1/2}\,\bpsi &\sqrt{i}\, \bpsi & 2\,\lambda^{1/2}
\end{array}
\right),\label{Ax}
\end{equation}\

Moreover, based on this fact it has been shown that the spatial component  of the B\"acklund transformation, and consequently the associated defect matrix, are also common  and henceforth universal 
within the entire  bosonic hierarchy \cite{Ana_proceedings, Ana15}.   This agrees  
More  recently, in \cite{1}, this result has been extended to the supersymmetric mKdV hierarchy  with the following defect matrix
\begin{equation}
\renewcommand{\arraystretch}{2}
	K=\left(
	\begin{array}{cc|c}
	\lambda^{1/2} & -\frac{2}{\omega^2}e^{\phi_+}\lambda^{-1/2} & -\frac{2\sqrt{i}}{\omega}e^{\frac{\phi_+}{2}}f_1 \hspace{0.2cm}\\
	-\frac{2}{\omega^2}e^{-\phi_+}\lambda^{1/2} & \lambda^{1/2} & -\frac{2\sqrt{i}}{\omega}e^{-\frac{\phi_+}{2}}f_1\lambda^{1/2}\\\hline
	\frac{2\sqrt{i}}{\omega}e^{-\frac{\phi_+}{2}}f_1\lambda^{1/2} & \frac{2\sqrt{i}}{\omega}e^{\frac{\phi_+}{2}}f_1 & \frac{2}{\omega^2}+\lambda^{1/2}
	\end{array}
	\label{Kmatrix}
	\right)
\end{equation}
where $ \phi_\pm =\phi_1\pm\phi_2$,  $\omega $ represents the B\"acklund parameter, and $f_1$ is an auxiliary fermionic field. \

We can now substitute \eqref{Ax} and \eqref{Kmatrix} in the $x$-part of the gauge transformation (\ref{gauge}), to get 
\begin{eqnarray}
	&& \pt_x\phi_-=\frac{4}{\omega^2}\sinh(\phi_+)-\frac{2i}{\omega}\sinh\left(\frac{\phi_+}{2}\right) f_1\bpp,\label{dxphi-}\\
	&& \bpsi_-= \frac{4}{\omega} \cosh\left(\frac{\phi_+}{2}\right)f_1,\label{f}\\
	&& \pt_xf_1=\frac{1}{\omega}\cosh\left(\frac{\phi_+}{2}\right)\bpp\label{dxf1}.
\end{eqnarray}
which is the spatial part of the B\"acklund transformations, where we have denoted $\bp_\pm=\bp_1\pm\bp_2 $.\

The corresponding temporal part of the B\"acklund is obtained from the  gauge transformation in (\ref{gauge}) for $\mu =0$,  i.e.,
\begin{equation}\label{gauget}
\pt_{t_{2n+1}}K=KA_{t_{2n+1}}(\phi_1,\bar{\psi}_1)-A_{t_{2n+1}}(\phi_2,\bar{\psi}_2)K.
\end{equation} 
where we consider the corresponding temporal part of the Lax pair $A_{t_{2n+1}}$. Now, we will consider some examples.  \

For $n=0$ we have that $A_x=A_{t_1}$ so the temporal part of the B\"acklund is, \
\br
\pt_{t_1}\phi_+&=&\pt_{x}\phi_+ ,\label{bt1ph+}\\
\pt_{t_1}f_1 &=& \pt_x f_1. \label{bt1f1}
\er

This implies that \
\br
\pt_{t_1}\phi_-=\pt_{x}\phi_-\label{bt1ph-}.
\er

Then, for $n=0$ the $x$ and $t$ components of the B\"acklund are the same.\

The next non-trivial example is the smKdV equation ($n=1$),
{\small
\begin{equation}
\newcommand\scalemath[2]{\scalebox{#1}{\mbox{\ensuremath{\displaystyle #2}}}}
\renewcommand{\arraystretch}{2.1}
 A_{t_3}=\left(
\scalemath{0.8}{\begin{array}{cc|c}
p_0 + \l^{1/2} p_{1/2}  -\l \dx+ \l^{3/2}  & p_+ -\l & \mu_+ +\l^{1/2} \nu_+ +\l \sqrt{i}\bpsi\\[0.1cm]
-\l p_- -\l^2& -p_0 +\l^{1/2} p_{1/2} +\l \dx +\l^{3/2}& \l^{1/2}\mu_- +\l \nu_- +\l^{3/2} \sqrt{i}\bpsi\\\hline
\l^{1/2}\mu_- -\l \nu_- +\l^{3/2} \sqrt{i}\bpsi& \mu_+ - \l^{1/2} \nu_++\l \sqrt{i}\bpsi & 2\l^{1/2} p_{1/2} +2 \l^{3/2}
\end{array}}
\right),\label{At3}
\end{equation}}
where \
\br
 p_0 &=&\frac{1}{4} \left(2(\dx)^3-\dxxx-3i\dx\bp \pt_x\bp \right), \, p_{1/2} = -\frac{i}{2}\bp\pa_x\bp,\,\nu_\pm = \frac{\sqrt{i}}{2}\big(\pa_x\bp \pm\bp \pa_x\phi\big)  \nonumber\\
 p_{\pm} &=& \frac{1}{2}\left(\dxx \pm (\dx)^2 \mp i\bp\pt_x\bp\right), \, \mu_\pm = \frac{\sqrt{i}}{4}\big(\pt^2_x\bp \pm\dx\pt_x\bp \mp \bp\dxx-2\bp(\dx)^2\big).\nonu\\
\er 

By substituting \eqref{At3} and \eqref{Kmatrix} in (\ref{gauget}), we obtain
\br 
4\pt_{t_3}\phi_-&=&\frac{i}{\o}\left[\phi^{(+)}_{2x}\cosh\Big(\frac{\phi_+}{2}\Big)-\left(\phi^{(+)}_{x}\right)^2\sinh\Big(\frac{\phi_+}{2}\Big)\]\bar{\psi}_+f_1-\frac{32}{\o^6}\,\sinh^3\phi_+ \nonu\\
 &\!\!\!\! -&\frac{i}{\o}\left[\phi^{(+)}_{x}\cosh\Big(\frac{\phi_+}{2}\Big)\bp^{(+)}_x - 2\sinh\Big(\frac{\phi_+}{2}\Big)\bp^{(+)}_{2x}\right]f_1\nonu\\
&\!\!\!\!   +&\frac{2}{\o^2}\Big[2\phi^{(+)}_{2x}\cosh\phi_+ -\left(\phi^{(+)}_{x}\right)^2\sinh\phi_+ +i\bar{\psi}_+\bar{\psi}^{(+)}_{x}\sinh\phi_+\Big]\nonu\\
& \!\!\!\! -&\frac{96i}{\o^5}\left[\sinh\Big(\frac{\phi_+}{2}\Big)+4\sinh^3\Big(\frac{\phi_+}{2}\Big) +3\sinh^5\Big(\frac{\phi_+}{2}\Big)\right]\bar{\psi}_+f_1,\label{dtphim}\\[0.1cm]
 4\pa_{t_3}f_1 &=& \frac{1}{2\o}\cosh\Big(\frac{\phi_+}{2}\Big) \left[2\dxxbpp -\bp_+ (\dxp)^2\]+\frac{12}{\o^5}\sinh^2\phi_+\cosh\Big(\frac{\phi_+}{2}\Big) \bp_+ \nonumber \\[0.1cm]
 &+&\frac{1}{2\o}\sinh\Big(\frac{\phi_+}{2}\Big) \left[\bp_+\dxxp -\dxp\dxbpp\]-\frac{12}{\o^4}\sinh\phi_+\cosh^2\Big(\frac{\phi_+}{2}\Big) \dxp f_1\nonu\\ \label{dtf1}.
\er
the corresponding temporal part of the B\"acklund transformation for the smKdV equation, where $\phi_{ix}^{(\pm)}=\pt^i_x\phi_{\pm}$ and  $\bp_{ix}^{(\pm)}=\pt^i_x\bp_{\pm}$. In \cite{1} this procedure have been aplied to obtain these transformations for the $t_5$ member of the smKdV hierarchy. 

\section{Recursion operator for the B\"acklund transformations}
In this section we will extend the idea of recursion operator to generate the B\"acklund transformation for smKdV hierarchy as an alternative method.\

In order to construct the recursion operator for the B\"acklund transformations we consider two different solutions of the equation (\ref{recursionphi}) as
\begin{eqnarray}
\frac{\pt \phi_1}{\pt t_{2n+3}}&=&\mathbb{R}_1^{(1)}\frac{\pt \phi_1}{\pt t_{2n+1}}+\mathbb{R}_2^{(1)}\frac{\pt \bp_1}{\pt t_{2n+1}},\qquad
\frac{\pt \bp_1}{\pt t_{2n+3}}=\mathbb{R}_3^{(1)}\frac{\pt \phi_1}{\pt t_{2n+1}}+\mathbb{R}_4^{(1)}\frac{\pt \bp_1}{\pt t_{2n+1}}\nonu\\ \label{phi1}\\
\frac{\pt \phi_2}{\pt t_{2n+3}}&=&\mathbb{R}_1^{(2)}\frac{\pt \phi_2}{\pt t_{2n+1}}+\mathbb{R}_2^{(2)}\frac{\pt \bp_2}{\pt t_{2n+1}},\qquad
\frac{\pt \bp_2}{\pt t_{2n+3}}=\mathbb{R}_3^{(2)}\frac{\pt \phi_2}{\pt t_{2n+1}}+\mathbb{R}_4^{(2)}\frac{\pt \bp_2}{\pt t_{2n+1}}\nonu\\ \label{phi2}
\end{eqnarray}
where $\mathbb{R}_i^{(p)}=\mathbb{R}_i\left(\dx^{(p)},\dxx^{(p)},\bp_p,\bp_x^{(p)},\bp_{2x}^{(p)}\right),\, i=1,...,4, \, p=1,2$.\

And take the following combination of these solutions \
\br
2\pt_{t_{2n+3}}\phi_-&=&\left(\mathbb{R}_1^{(1)}+\mathbb{R}_1^{(2)}\right)\pt_{t_{2n+1}}\phi_-+\left(\mathbb{R}_2^{(1)}+\mathbb{R}_2^{(2)}\right)\pt_{t_{2n+1}}\bp_-\nonu\\&+&\left(\mathbb{R}_1^{(1)}-\mathbb{R}_1^{(2)}\right)\pt_{t_{2n+1}}\phi_++\left(\mathbb{R}_2^{(1)}-\mathbb{R}_2^{(2)}\right)\pt_{t_{2n+1}}\bp_+ ,\label{phim}
\er
\begin{eqnarray}
2\pt_{t_{2n+3}}\bp_-&=&\left(\mathbb{R}_3^{(1)}+\mathbb{R}_3^{(2)}\right)\pt_{t_{2n+1}}\phi_-+\left(\mathbb{R}_4^{(1)}+\mathbb{R}_4^{(2)}\right)\pt_{t_{2n+1}}\bp_-\nonu\\&+&\left(\mathbb{R}_3^{(1)}-\mathbb{R}_3^{(2)}\right)\pt_{t_{2n+1}}\phi_++\left(\mathbb{R}_4^{(1)}-\mathbb{R}_4^{(2)}\right)\pt_{t_{2n+1}}\bp_+ \label{psim}.
\end{eqnarray} 
%where we introduced the new variables $\phi_{\pm}=\phi_1\pm\phi_2$ and $\bp_{\pm}=\bp_1\pm\bp_2$. 

At this point, we conjecture that the equations (\ref{phim}) and (\ref{psim}) correspond to the temporal part of the super B\"acklund transformation for an super integrable equation especified by $n$. We note that 
as well as the consecutive equations of motion within the hierarchy are connected by the same recursion operator, here the same occurs to the B\"acklund transformations.  In order to clarify this hypothesis we next
consider  some examples.\

For $n=0$ we have\
\begin{eqnarray}
2\pt_{t_{3}}\phi_-&=&\left(\mathbb{R}_1^{(1)}+\mathbb{R}_1^{(2)}\right)\pt_{t_{1}}\phi_-+\left(\mathbb{R}_2^{(1)}+\mathbb{R}_2^{(2)}\right)\pt_{t_{1}}\bp_-\nonu\\&+&\left(\mathbb{R}_1^{(1)}-\mathbb{R}_1^{(2)}\right)\pt_{t_{1}}\phi_++\left(\mathbb{R}_2^{(1)}-\mathbb{R}_2^{(2)}\right)\pt_{t_{1}}\bp_+ ,\\
2\pt_{t_{3}}\bp_-&=&\left(\mathbb{R}_3^{(1)}+\mathbb{R}_3^{(2)}\right)\pt_{t_{1}}\phi_-+\left(\mathbb{R}_4^{(1)}+\mathbb{R}_4^{(2)}\right)\pt_{t_{1}}\bp_-\nonu\\&+&\left(\mathbb{R}_3^{(1)}-\mathbb{R}_3^{(2)}\right)\pt_{t_{1}}\phi_++\left(\mathbb{R}_4^{(1)}-\mathbb{R}_4^{(2)}\right)\pt_{t_{1}}\bp_+ .
\end{eqnarray} 

By using (\ref{bt1ph+})-(\ref{bt1ph-}) we recover equations (\ref{dtphim}) and (\ref{dtf1}), ie the time component of the B\"acklund transformantions for $n=2$ (smKdV). \ 

Next we consider the case for $n=2$ and using again (\ref{dtphim})-(\ref{dtf1}) we obtain from (\ref{phim}),\
\begin{eqnarray}
16\pt_{t_5}\phi_-&=&\dxxxxxm+\frac{3}{8}\left(\dxm\right)^5-\frac{5}{2}\dxm\left(\left(\dxxm\right)^2+\left(\dxxp\right)^2\right)-5\dxxm\dxxp\dxp\nonu\\
&+&5\dxm\dxp\left(\frac{3}{8}\left(\dxp\right)^3-\dxxxp\right)-\frac{5}{2}\dxxxm\left(\left(\dxm\right)^2+\left(\dxp\right)^2\right)\nonu\\
&+&\frac{5i}{4}\left(\bpm\dxbpm+\bpp\dxbpp\right)\left[\dxxxm-\dxm\left(\left(\dxm\right)^2+3\left(\dxp\right)^2\right)\right]\nonu\\
&+&\frac{5i}{4}\left(\bpm\dxbpp+\bpp\dxbpm\right)\left[\dxxxp-\dxp\left(\left(\dxp\right)^2+3\left(\dxm\right)^2\right)\right]\nonu\\
&+&\frac{5i}{4}\bpp\left(\dxxxbpm\dxp+\dxxxbpp\dxm\right)+\frac{5i}{4}\bpp\left(\dxxbpm\dxxp+\dxxbpp\dxxm\right)\nonu\\
&+&\frac{15}{4}\left(\dxm\right)^3\left(\dxp\right)^2+\frac{5i}{4}\bpm\left(\dxxm\dxxbpm+\dxxp\dxxbpp\right)\nonu\\
&+&\frac{5i}{4}\bpm\left(\dxm\dxxxbpm+\dxp\dxxxbpp\right)
\end{eqnarray}

And for the equation (\ref{psim}) we get\
\begin{eqnarray}
16\pt_{t_5}\bp_-&=&\dxxxxxbpm-\frac{5}{4}\bpm\left(\dxm\dxxxxm+\dxp\dxxxxp\right)-\frac{5}{4}\bpp\left(\dxm\dxxxxp+\dxp\dxxxxm\right)\nonu\\
&-&\frac{5}{4}\left(\bpm\dxxp+\bpp\dxxm\right)\left[2\dxxxp-\dxp\left(\left(\dxp\right)^2+3\left(\dxm\right)^2\right)\right]\nonu\\
&-&\frac{5}{4}\left(\bpm\dxxm+\bpp\dxxp\right)\left[2\dxxxm-\dxm\left(\left(\dxm\right)^2+3\left(\dxp\right)^2\right)\right]\nonu\\
&-&\frac{5}{8}\dxbpm\dxm\left[6\dxxxm-\dxm\left(\left(\dxm\right)^2+6\left(\dxp\right)^2\right)\right]\nonu\\
&-&\frac{5}{8}\dxbpm\dxp\left(6\dxxxp-\left(\dxp\right)^3\right)-\frac{5}{4}\dxbpp\left(4\dxxm\dxxp+3\dxxxp\dx\right)\nonu\\
&-&\frac{5}{4}\dxbpp\dxp\left[3\dxxxm-2\left(\left(\dxm\right)^2+\left(\dxp\right)^2\right)\right]\nonu\\
&-&\frac{15}{4}\dxxbpp\left(\dxxp\dxm+\dxxm\dxp\right)-\frac{5}{4}\dxxxbpm\left(\left(\dxm\right)^2+\left(\dxp\right)^2\right)\nonu\\
&-&\frac{5}{2}\dxxxbpp\dxm\dxp-\frac{5}{2}\dxbpm\left(\left(\dxxm\right)^2+\left(\dxxp\right)^2\right)\nonu\\
&-&\frac{15}{4}\dxxbpm\left(\dxxm\dxm+\dxxp\dxp\right)
\end{eqnarray}

Now, using the $x$-part of the B\"acklund transformation (\ref{dxphi-})-(\ref{dxf1}) in these two equations we end up with the corresponding B\"acklund transformation for $n=3$, that was obtained in \cite{1}. 

\section*{Conclusions}

In this note we have considered a hierarchy of supersymmetric equations of motion  underlined  by an affine construction of a Kac-Moody algebra $\hat {sl}(2,1)$.
These equations of motion  were shown to be related by a recursion operator  that maps  consecutive  time flows.
Moreover, it was shown that   the B\"acklund transformation  follows the same  relation  generated by the recursion operator.  Such framework provides a general and systematic
method of   constructing  B\"acklund transformations  for the entire hierarchy completing and clarifying the  question raised in ref. \cite{1}.\

An interesting point  we  would like to point out  and is still  under investigation  is about the construction  of the underlying affine Kac-Moody algebra.
The key ingredient  in the affinization   was the  decomposition of the integers and semi integers numbers according to a 
$Z_2$ structure assigned to    bosonic and fermionic generators  defined in (\ref{affine}).
For higher rank algebras we expect to systematize such construction decomposing  both integers and semi-integers in disjoint subsets   compatible with the closure of the algebra,  e.g. $Z_k$ for $\hat {sl}(k,1)$.

\begin{acknowledgement}
ALR thanks the Sao Paulo Research Foundation FAPESP for financial support under the process 2015/00025-9. JFG, NIS and AHZ thank CNPq for financial support.
\end{acknowledgement}
\section*{Appendix}
\addcontentsline{toc}{section}{Appendix}\label{appA}
Here we resume the commutation and Anti-commutation relations for the $ \widehat{\textit{sl}} $(2,1) affine Lie superalgebra

\br
&&\lbrack K_{1}^{(2n+1)},K_{2}^{(2m+1)}\rbrack =0,\nonumber\\
&&\lbrack M_{1}^{(2n+1)},K_{1}^{(2m+1)}\rbrack=2M_{2}^{2(n+m+1)}+(n+m)\delta_{n+m+1,0}\hat{c},\nonumber\\
&&\lbrack M_{1}^{(2n+1)},K_{2}^{(2m+1)}\rbrack=0,\nonumber\\
&&\lbrack K_{2}^{(2n+1)},K_{2}^{(2m+1)}\rbrack=-(n-m)\delta_{n+m+1,0}\hat{c},\nonumber\\
&&\lbrack M_{2}^{(2n)},K_{1}^{(2m+1)}\rbrack=2M_{1}^{2(n+m)+1},\nonumber\\
&&\lbrack M_{2}^{(2n)},K_{2}^{(2m+1)}\rbrack=0,\nonumber\\
&&\lbrack M_{1}^{(2n+1)},M_{2}^{(2m)}\rbrack=-2K_{1}^{2(n+m)+1},\nonumber\\
&&\lbrack M_{1}^{(2n+1)},M_{1}^{(2m+1)}\rbrack=-(n-m)\delta_{n+m+1,0}\hat{c},\nonumber\\
&&\lbrack M_{2}^{(2n)},M_{2}^{(2m)}\rbrack=(n-m)\delta_{n+m,0}\hat{c},\nonumber\\
&&\lbrack K_{1}^{(2n+1)},K_{1}^{(2m+1)}\rbrack=(n-m)\delta_{n+m+1,0}\hat{c},\nonu\\
&&\lbrack F_{1}^{(2n+3/2)},K_{1}^{(2m+1)}\rbrack=-\lbrack F_{1}^{(2n+3/2)},K_{2}^{(2m+1)}\rbrack=F_{2}^{2(n+m+1)+1/2},\nonumber\\
&&\lbrack F_{2}^{(2n+1/2)},K_{1}^{(2m+1)}\rbrack=-\lbrack F_{2}^{(2n+1/2)},K_{2}^{(2m+1)}\rbrack=F_{1}^{2(n+m)+3/2},\nonumber\\
&&\lbrack M_{1}^{(2n+1)},F_{1}^{(2m+3/2)}\rbrack=G_{1}^{2(n+m+1)+1/2},\nonumber\\
&&\lbrack M_{1}^{(2n+1)},F_{2}^{(2m+1/2)}\rbrack=-\lbrack M_{2}^{(2n)},F_{1}^{(2m+3/2)}\rbrack=G_{2}^{2(n+m)+3/2},\nonumber\\
&&\lbrack M_{2}^{(2n)},F_{2}^{(2m+1/2)}\rbrack=-G_{1}^{2(n+m)+1/2},\nonumber\\
&&\lbrack G_{1}^{(2n+1/2)},K_{1}^{(2m+1)}\rbrack=-G_{2}^{2(n+m)+3/2},\nonumber\\
&&\lbrack G_{1}^{(2n+1/2)},K_{2}^{(2m+1)}\rbrack=-G_{2}^{2(n+m)+3/2},\nonumber\\
&&\lbrack G_{2}^{(2n+3/2)},K_{1}^{(2m+1)}\rbrack=-G_{1}^{2(n+m+1)+1/2},\nonumber\\
&&\lbrack G_{2}^{(2n+3/2)},K_{2}^{(2m+1)}\rbrack=-G_{1}^{2(n+m+1)+1/2},\nonumber\\
&&\lbrack M_{1}^{(2n+1)},G_{1}^{(2m+1/2)}\rbrack=-F_{1}^{2(n+m)+3/2},\nonumber\\
&&\lbrack M_{1}^{(2n+1)},G_{2}^{(2m+3/2)}\rbrack=-F_{2}^{2(n+m+1)+1/2},\nonumber\\
&&\lbrack M_{2}^{(2n)},G_{1}^{(2m+1/2)}\rbrack=-F_{2}^{2(n+m)+1/2},\nonumber\\
&&\lbrack M_{2}^{(2n)},G_{2}^{(2m+3/2)}\rbrack=-F_{1}^{2(n+m)+3/2},\nonumber\\
&&\{F_{1}^{(2n+3/2)},F_{2}^{(2m+1/2)}\}=[(2n+1)-2m]\delta_{n+m+1,0}\hat{c},\nonumber
\er
\begin{eqnarray}
&&\{F_{1}^{(2n+3/2)},F_{1}^{(2m+3/2)}\}=2(K_{2}^{2(n+m+1)+1}+K_{1}^{2(n+m+1)+1}),\nonumber\\
&&\{F_{2}^{(2n+1/2)},F_{2}^{(2m+1/2)}\}=-2(K_{2}^{2(n+m)+1}+K_{1}^{2(n+m)+1}),\nonumber\\
&&\{F_{2}^{(2n+1/2)},G_{1}^{(2m+1/2)}\}=2M_{1}^{2(n+m)+1},\nonumber\\
&&\{F_{1}^{(2n+3/2)},G_{2}^{(2m+3/2)}\}=-2M_{1}^{2(n+m+1)+1},\nonumber\\
&&\{F_{1}^{(2n+3/2)},G_{1}^{(2m+1/2)}\}=2M_{2}^{2(n+m+1)}+[(2n+1)+2m]\delta_{n+m+1,0}\hat{c},\nonumber\\
&&\{F_{2}^{(2n+1/2)},G_{2}^{(2m+3/2)}\}=-2M_{2}^{2(n+m+1)}-[2n+(2m+1)]\delta_{n+m+1,0}\hat{c},\nonumber\\
&&\{G_{1}^{(2n+1/2)},G_{2}^{(2m+3/2)}\}=[2n-(2m+1)]\delta_{n+m+1,0}\hat{c},\nonumber\\
&&\{G_{1}^{(2n+1/2)},G_{1}^{(2m+1/2)}\}=2(K_{2}^{2(n+m)+1}-K_{1}^{2(n+m)+1}),\nonumber\\
&&\{G_{2}^{(2n+3/2)},G_{2}^{(2m+3/2)}\}=-2(K_{2}^{2(n+m+1)+1}-K_{1}^{2(n+m+1)+1}).
\end{eqnarray}

\end{document}